Original Paper

# Medical Misinformation in AI-Assisted Self-Diagnosis: Development of a Method (EvalPrompt) for Analyzing Large Language Models

Troy Zada[1*], BASc; Natalie Tam[1*], BASc; Francois Barnard[1], BASc; Marlize Van Sittert[2], BS; Venkat Bhat[3,4], MSc, MD; Sirisha Rambhatla[1], PhD

[1]Department of Management Sciences and Engineering, University of Waterloo, Waterloo, ON, Canada
[2]Faculty of Law, University of Toronto, Toronto, ON, Canada
[3]Department of Psychiatry, University of Toronto, Toronto, ON, Canada
[4]Interventional Psychiatry Program, St. Michael's Hospital, Unity Health Toronto, Toronto, ON, Canada
*these authors contributed equally

**Corresponding Author:**
Sirisha Rambhatla, PhD
Department of Management Sciences and Engineering
University of Waterloo
200 University Avenue West
Waterloo, ON, N2L 3G1
Canada
Phone: 1 5198884567 ext 33279
Email: sirisha.rambhatla@uwaterloo.ca

## Abstract

**Background:** Rapid integration of large language models (LLMs) in health care is sparking global discussion about their potential to revolutionize health care quality and accessibility. At a time when improving health care quality and access remains a critical concern for countries worldwide, the ability of these models to pass medical examinations is often cited as a reason to use them for medical training and diagnosis. However, the impact of their inevitable use as a self-diagnostic tool and their role in spreading health care misinformation has not been evaluated.

**Objective:** This study aims to assess the effectiveness of LLMs, particularly ChatGPT, from the perspective of an individual self-diagnosing to better understand the clarity, correctness, and robustness of the models.

**Methods:** We propose the comprehensive testing methodology evaluation of LLM prompts (EvalPrompt). This evaluation methodology uses multiple-choice medical licensing examination questions to evaluate LLM responses. Experiment 1 prompts ChatGPT with open-ended questions to mimic real-world self-diagnosis use cases, and experiment 2 performs sentence dropout on the correct responses from experiment 1 to mimic self-diagnosis with missing information. Humans then assess the responses returned by ChatGPT for both experiments to evaluate the clarity, correctness, and robustness of ChatGPT.

**Results:** In experiment 1, we found that ChatGPT-4.0 was deemed correct for 31% (29/94) of the questions by both nonexperts and experts, with only 34% (32/94) agreement between the 2 groups. Similarly, in experiment 2, which assessed robustness, 61% (92/152) of the responses continued to be categorized as correct by all assessors. As a result, in comparison to a passing threshold of 60%, ChatGPT-4.0 is considered incorrect and unclear, though robust. This indicates that sole reliance on ChatGPT-4.0 for self-diagnosis could increase the risk of individuals being misinformed.

**Conclusions:** The results highlight the modest capabilities of LLMs, as their responses are often unclear and inaccurate. Any medical advice provided by LLMs should be cautiously approached due to the significant risk of misinformation. However, evidence suggests that LLMs are steadily improving and could potentially play a role in health care systems in the future. To address the issue of medical misinformation, there is a pressing need for the development of a comprehensive self-diagnosis dataset. This dataset could enhance the reliability of LLMs in medical applications by featuring more realistic prompt styles with minimal information across a broader range of medical fields.

**Keywords:** ChatGPT; health care; LLM; misinformation; self-diagnosis; large language model

# Introduction

## Background

Large language models (LLMs) have grown in popularity with an ever-expanding list of applications due to their efficiency and accessibility [1,2]. With their increased usage, LLMs are gaining user trust [3], partly due to the anthropomorphic responses produced by models such as GPT-4o, even though they can generate misinformation at scale [4,5]. Recent reports highlight the inability of differentiating truth from misinformation, and the potential collapse of health care systems, as major disruptors on the horizon [6]. This emphasizes the urgent need to develop solutions to ensure the delivery of factual information.

In health care, self-diagnosis through web searches has become a widespread practice and is especially important for underserved communities [7-10], which means that the prospective usage of LLMs in this domain is inevitable. However, relying solely on online searches for health information can result in severe misinformation as content on social media often spreads more rapidly than scientific knowledge [11]. Inaccurate content, conspiracy theories, and false claims are all forms of misinformation which can impact public perceptions, alter behaviors, and reduce trust in health care systems [12]. Moreover, the ongoing global shortage of health care workers [13-16] has driven government entities and health care organizations to explore the use of LLMs as health care assistants and expertise replacements for diagnosis and education [17-23]. Thus, there is a need to study the quality and reliability of LLM-generated responses to health care–related questions.

## Prior Work

Recent work has focused on analyzing ChatGPT across various industries, including its application within health care. ChatGPT is a natural language processing model distinctive for its narrative response style to user input [24,25]. Studies have assessed its performance on examinations [20,24,26] and its utility as a self-studying tool [24,27,28], leveraging its ability to provide tailored responses and immediate feedback. Furthermore, ChatGPT has demonstrated potential in assisting research and academic writing by enhancing efficiency and mitigating gaps in researcher knowledge [29]. However, the increased usage of ChatGPT raises significant ethical concerns regarding plagiarism, bias, transparency, inaccuracy, and health equity [30-34].

ChatGPT has also demonstrated superior performance in answering medical questions compared to other LLMs [35]. Research has explored its applications in medical education, including its effectiveness on licensing examinations, tailored learning experiences, and comprehension of complex medical concepts and clinical reasoning [20,24,27,33,36]. Other areas of study have focused on identifying inefficiencies and inaccuracies within clinical workflows, medical research, and diagnoses, with the objective of integrating LLMs to optimize documentation, triage, and clinical data management procedures [33,37-44]. Additionally, investigations into diagnostic assistance have integrated patient questionnaires and medical imaging with LLMs [43,45-48]. Despite the apparent high performance of LLMs in health care, they remain inferior compared to the judgment of human clinicians [49].

In summary, prior work has generally evaluated the trustworthiness of LLMs [50], along with specifically examining their performance in medical situations. These methods have used the multiple-choice questions from the United States Medical Licensing Exam (USMLE) [51] to evaluate the capability of LLMs in attaining scores near the passing threshold of 60% [20,52]. However, these evaluations inaccurately depict the capability of LLMs for self-diagnosis. In practical situations where individuals use technology to self-diagnose, they would not include answers when posing questions and would not provide the same level of information as the examination questions.

## Goal of This Study

This study critically examines the performance of LLMs in responding to health care–related questions. To achieve this, we propose evaluation of LLM prompts. This evaluation procedure contains detailed guidelines to assess ChatGPT's response to open-ended questions and validate the robustness of these responses using a sentence dropout method. This 2-staged approach, to our knowledge, is the first comprehensive strategy aimed at better understanding LLM responses and their implications for medical misinformation. We hypothesize that LLMs, particularly ChatGPT-4.0 (referred to as GPT-4.0), are currently unsuitable for self-diagnosis purposes since a significant portion of responses will be ambiguous or incorrect. In particular, this hypothesis is validated if GPT-4.0 surpasses a minimum threshold of 60% [20] for each of the following three questions:

1. Are the responses clear? This question can be answered by analyzing the response consistency.
2. Are the responses genuinely correct? This question can be answered by identifying the responses classified as correct by all assessors.
3. Are the responses robust? This question can be answered by conducting an ablation study on the correct responses.

# Methods

## Study Design

In this section, the considerations and preparations for the decided dataset are first specified. Then, the assessor procedures and guidelines are discussed along with the process used to analyze the output answers from ChatGPT. Finally, the complete testing methodology is introduced. The overall process uses both nonmedical and medical experts for assessment and is segmented into two experiments: (1) ChatGPT responses on USMLE step 1 open-ended prompts and (2) ChatGPT robustness and ablation study analysis. The GPT-4.0 responses were generated for each of the USMLE questions using Python, with the code and datasets available in the EvalPrompt (evaluation of large language model prompts) repository [53].

## Dataset Considerations and Preparations

### Overview

The USMLE [51] dataset consists of 3 steps undertaken by medical students throughout their program. Each step is a test consisting of multiple-choice, single-answer, and no-justification questions. Particularly, in this work, the questions are extracted from step 1 since the step 2 and step 3 questions are medically complex for the general population. Furthermore, only textual questions were kept, resulting in a dataset containing 94 single-answer questions that would be used for prompting ChatGPT. Textbox 1 [51] displays a sample of an extracted step 1 question.

**Textbox 1.** Sample question 3.1 is directly extracted from the United States Medical Licensing Exam step 1 test. This question is a multiple-choice question with a correct answer of (D).

> Question: In a sample of 100 individuals, the mean leukocyte count is 7500/mm$^3$, with a standard deviation of 1000/mm$^3$. If the leukocyte counts in this population follow a normal (gaussian) distribution, approximately 50% of individuals will have which of the following total leukocyte counts?
> (A) 5500–9500/mm$^3$
> (B) <6500/mm$^3$ or >8500/mm$^3$
> (C) 6500–8500/mm$^3$
> (D) <7500/mm$^3$
> (E) >9500/mm$^3$
> Answer: (D) <7500/mm$^3$

### Baseline ChatGPT Answer Analysis

The first experiment established the foundation for all testing. From the initial multiple-choice questions, each question was transformed into an open-ended question to accurately simulate the circumstances of an individual interacting with ChatGPT. This process was accomplished by removing the multiple-choice options and replacing any instance of "which of the following" with "what." An example transformation is provided in Textbox 2 [51], where the original question from Textbox 1 was transformed into an open-ended question.

**Textbox 2.** Transformed United States Medical Licensing Exam step 1 test question based on the original Textbox 1 question. The transformation involves removing the (A)-(E) options and replacing the text "which of the following" with "what."

> Question: In a sample of 100 individuals, the mean leukocyte count is 7500/mm$^3$, with a standard deviation of 1000/mm$^3$. If the leukocyte counts in this population follow a normal (gaussian) distribution, approximately 50% of individuals will have what total leukocyte counts?

### ChatGPT Robustness and Ablation Study Analysis

The second experiment was built upon the foundation established in the first experiment by using only the questions that were categorized as correct (C) by all assessors. Using this subset of data, new questions were iteratively generated by removing a sentence from the question and prompting ChatGPT with the modified question. This process is repeated across all the sentences of a question, excluding the final question sentence indicated by the "what...?" structure. For instance, if a question has 3 sentences, then 3 versions of the same question would be generated, where each question variation has 2 sentences. The first version would remove the first sentence, the second version would remove the second sentence, and the third version would remove the third sentence. The ablation study aimed to simulate how a self-diagnosing individual may overlook information within their health-related searches.

Each of the open-ended modified questions was then provided to ChatGPT using the OpenAI application programming interface to aggregate the LLM responses for the questions. Specifically, the *gpt-4* model was used for GPT-4.0, along with specific parameters to ensure consistency among the responses and that each question would be answered independently. For instance, the model's temperature was set to zero to ensure the responses were focused and deterministic. Details of the complete GPT-4.0 model and corresponding parameters are provided in Multimedia Appendix 1, along with Textbox 3 [51] displaying the GPT-4.0 response to the open-ended Textbox 2 question.

**Textbox 3.** GPT-4.0 response to United States Medical Licensing Exam step 1 test question based on the transformed Textbox 2 question. The assessor can then compare the GPT-4.0 answer with the original multiple-choice options to evaluate the correctness of the response.

> Answer: Approximately 50% of individuals will have total leukocyte counts of 7500/mm³ or less. Explanation: In a normal distribution, the mean value is the point at which approximately 50% of the data falls below and 50% falls above. Therefore, in this case, where the mean leukocyte count is 7500/mm³, approximately 50% of individuals will have leukocyte counts of 7500/mm³ or less.

## Assessor Procedures and Guidelines

The 94 open-ended questions served as prompts for ChatGPT, where 3 nonexperts and 3 experts evaluated the generated responses. Nonexperts were used to simulate the self-diagnosis process, while experts were used to assess the response quality.

Each assessor was provided with detailed procedures and guidelines as outlined in Multimedia Appendix 2 to ensure the response categorization process was standardized among the assessors. Assessors were instructed to categorize the responses as either C, partially correct (PC), incorrect (I), or ambiguous (A), and were provided with detailed examples to distinguish between the categories. Moreover, assessors were directed to evaluate the responses independently to ensure no bias.

## Testing Methodology

We develop the testing methodology EvalPrompt, as summarized in Figure 1. In total, $N_q = 94$ questions were selected from the USMLE step 1 test. Each question contained a question portion $x_j$, answer options $o_j$, and a correct answer $y_j$, altogether forming the dataset of multiple-choice questions $D = \{x_j, o_j, y_j\}_{j=1}^{N_q}$. Subsequently, each question, $x_j$, was extracted and transformed into an open-ended variation which was then presented to ChatGPT.

**Figure 1.** EvalPrompt summary. (1) A subset of 94 USMLE step 1 [51] questions consisting of multiple-choice, single-answer questions are selected. (2) These 94 questions are modified to produce open-ended prompts. (3) The open-ended prompts are processed through the ChatGPT API (4) ChatGPT's answers are recorded and presented to $k$ independent assessors to categorize as either correct, partially correct, incorrect, or ambiguous. (5) The categorizations classified as correct by all assessors are aggregated to formulate a new dataset for sensitivity analysis. (6) The prompts produced from the iterative sentence dropout are processed through ChatGPT. (7) The independent assessors categorize ChatGPT's answers. These categorizations are then aggregated based on the agreement where all assessors categorized the answer as correct. The resulting dataset, $W$, is the subset of USMLE question variations that ChatGPT answered correctly. USMLE: United States Medical Licensing Exam.

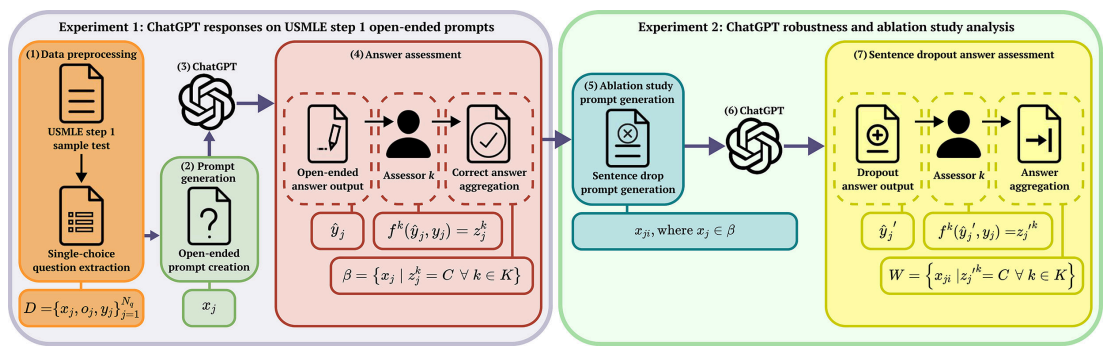

After collecting and processing the answers from ChatGPT, 3 nonexpert and 3 expert assessors independently evaluated each ChatGPT answer, $\hat{y}_j$. In total, $K = 6$ assessors were used. The assessors were instructed to provide a label $z \in Z; Z = \{C, PC, I, A\}$ denoting whether the answer, in comparison to the ground truth, $y_j$, was C, PC, I, or A, respectively. The assessment process can be expressed as a function $f^k(\hat{y}_j, y_j) = z_j^k$ for the $k^{th}$ assessor, where the assessor would use the correct answer to the question along with ChatGPT's answer to categorize the response.

After the $k$ assessors finished the categorization process, the questions categorized as C by all $k$ assessors were aggregated to define a subset dataset, $\beta$. Once the resulting dataset, $\beta = \{x_j | z_j^k = C \forall k \in K\}$ was generated, the first experiment was concluded.

The second experiment was built upon the dataset and assessments completed during the first experiment. An ablation study was conducted over the questions within dataset $\beta$ via an iterative sentence dropout by performing $i - 1$ iterations over an open-ended prompt $x_j$, where $i$ is the number of sentences in the prompt. For the $i^{th}$ iteration, the $i^{th}$ sentence was removed from the prompt before running the prompt through ChatGPT, as expressed by $x_{ji}$, where $x_j \in \beta$. Each $i^{th}$ sentence was iteratively removed and processed except for the final sentence which contained the question sentence. The final sentence of the question was mandatory to include to ensure ChatGPT provided an appropriate response.

After processing the sentence dropout questions through ChatGPT, the same 6 assessors evaluated the responses, $\hat{y}_j'$. The assessment process can again be expressed as a function $f^k(\hat{y}_j, y_j) = z_j^{`k}$ for the $k^{th}$ assessor, where the assessor would use the correct answer to the question along with ChatGPT's answer to categorize the response. This generated the resulting dataset, $W$, defined as $W = \{x_{ji} | z_j^{`k} = C \forall k \in K\}$, being the subset of USMLE question variations for the certainly correctly prompts.

## Ethical Considerations

Since the aim of this study is to analyze ChatGPT and not human subjects, a research ethics board review was not required. The evaluation of ChatGPT's responses was carried out on a volunteer basis, and all assessors were informed that contributing to the experiments would not result in any safety or privacy risks.

# Results

## Are the Responses Clear?

To determine if the GPT-4.0 responses are clear, response categorizations across assessors were compared. As displayed in Figure 2, categorization disparities exist across the nonexpert and expert assessors with few alignments across assessors as depicted in Table 1. Namely, the number of C categorizations ranges from 39 to 51, while the number of I categorizations ranges from 12 to 39. This wide range of values for the categories immediately depicts the uncertainty across assessors due to each assessor having varying backgrounds and levels of medical expertise.

Analyzing the categorization for each question independently, many discrepancies exist among the assessors. As detailed in Table 2, on average 52% ([(52+46)/2]/94, SD of 3) of the responses in experiment 1 and 73% ([(119+104)/2]/152, SD of 7.5) of the responses in experiment 2 were categorized identically by the assessors. For example, USMLE question 54.1 was categorized by the nonexperts as PC, A, and C, and by the experts as I, PC, and PC, respectively. The inconsistency in categorizations depicts that LLM responses do not yet have a single apparent answer, but rather are still subject to interpretation depending on the individual.

**Figure 2.** Experiment 1: ChatGPT responses on USMLE step 1 open-ended prompts. Individual categorizations for the nonexpert and expert assessors, where each bar represents an individual's categorizations. The categorizations drastically vary across assessors, where some assessors categorized more responses as correct while other assessors categorized more responses as incorrect on the same dataset. USMLE: United States Medical Licensing Exam.

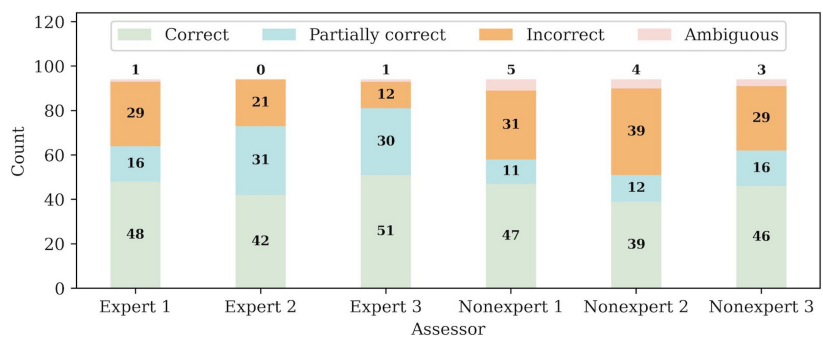

**Table 1.** Experiment 1: ChatGPT responses on USMLE[a] step 1 open-ended prompts. Overall response categorization between nonexpert and expert assessors. The top left section indicates the number of instances, regardless of correctness, where all nonexpert and expert assessors categorized a question identically. The bottom right section indicates the total number of questions categorized as correct by all nonexpert and expert assessors.

|  | Expert | | | |
|---|---|---|---|---|
|  | Identical | Mismatch | Correct | Other |
| Nonexpert | | | | |
|   Identical | 32 | 20 | —[b] | — |
|   Mismatch | 14 | 28 | — | — |
|   Correct | — | — | 29 | 4 |
|   Other | — | — | 6 | 55 |

[a]USMLE: United States Medical Licensing Exam.
[b]Not applicable.

**Table 2.** Number of identical categorizations among all individuals within the 2 groups for the 2 experiments. Experiment 1 had 94 questions in total, with 52 and 46 of the questions being categorized the same among the nonexpert and expert assessors, respectively. Similarly, experiment 2 had 152 responses in total spanning 29 unique questions, with 119 and 104 of the questions being categorized the same among the nonexpert and expert assessors, respectively.

|  | Experiment 1 | | Experiment 2 | |
| --- | --- | --- | --- | --- |
|  | Nonexpert | Expert | Nonexpert | Expert |
| Correct | 33 | 35 | 111 | 96 |
| Partially correct | 1 | 4 | 0 | 1 |
| Incorrect | 18 | 7 | 8 | 7 |
| Ambiguous | 0 | 0 | 0 | 0 |
| Overall agreement | 52 | 46 | 119 | 104 |

The nonexpert and expert assessors reached an identical conclusion for 52 and 46 of the questions for experiment 1, respectively, as detailed in Table 2. However, the groups collectively could only reach an identical conclusion for 32 of the questions as listed in Table 1. Since only 34% (32/94) of the responses were consistent across all assessors, the 60% threshold could not be met indicating that the responses are unclear. The decrease in the collective number of identically classified responses suggests that the GPT-4.0 responses are still too ambiguous for assessors to reach appropriate conclusions. The responses are not obvious enough such that anyone, regardless of their background and expertise, can reach the same conclusion.

### Are the Responses Genuinely Correct?

To determine the number of GPT-4.0 responses that are genuinely correct, the responses categorized as C by all nonexpert and expert assessors were analyzed and compared. As provided in Table 1, the experiment 1 categorizations where all assessors agreed are listed. In particular, 29 of the 94 responses were categorized as C, meaning that GPT-4.0 is certainly correct 31% (29/94) of the time.

The limited number of C responses indicates that GPT-4.0 is not often factual. As portrayed in Table 2, nonexpert and expert assessors classified 33 and 35 of the responses as C, respectively, even though collectively 29 responses were considered C. Since only 31% (29/94) of the responses were considered C across all assessors, the 60% threshold could not be met indicating that the responses are mostly incorrect. Hence, even though both groups approximately categorized the same number of responses as C, there are still many responses that are either A or I.

### Are the Responses Robust?

To determine if the GPT-4.0 responses are robust, an ablation study was conducted on the 29 correct responses. Reprompting GPT-4.0 with similar variations of the correct questions tested its ability to attain the correct answer even with information missing. This process aimed to simulate the self-diagnosis process since each individual would prompt LLMs with varying levels of information; some individuals would provide extensive details, while others may provide limited information. Thus, assessing GPT-4.0's robustness.

Table 2 provided the categorization details of the ablation study, consisting of 29 unique questions with 152 question variations. On average 68% ([(111+96)/2]/152, SD of 7.5) of the responses continue to be categorized as C even after removing information for each group. These results are far greater than GPT-4.0's accuracy on the initial 94 questions, which was correct only 31% (29/94) of the time. Moreover, Figure 3 depicts the categorizations for each of the 6 assessors. Assessors categorized the experiment 2 dataset C much more frequently than the experiment 1 dataset. On average, 80% ([(127+120+105+116+133+129)/6]/152, SD of 9.34) of the assessors categorized the questions as C, depicting GPT-4.0's robustness on answers that are certainly correct.

Table 3 also displays the assessment similarities between the nonexpert and expert assessors for experiment 2. A total of 92 of the 152 questions were categorized C by all nonexpert and expert assessors, meaning that 61% (92/152) of the responses are certainly correct. Since 61% (92/152) of the responses for the sentence dropout experiment were categorized as C across all assessors, the 60% threshold is met indicating that the responses are robust. In other words, if GPT-4.0 correctly answers a question, it is likely to correctly answer a similar variation of the question again, even if some information is missing.

**Figure 3.** Experiment 2: ChatGPT robustness and ablation study analysis. Individual categorization for the nonexpert and expert assessors, where each bar represents an individual's categorizations. Out of the 152 questions, the assessors on average categorized 122 of the questions as correct, depicting its robustness.

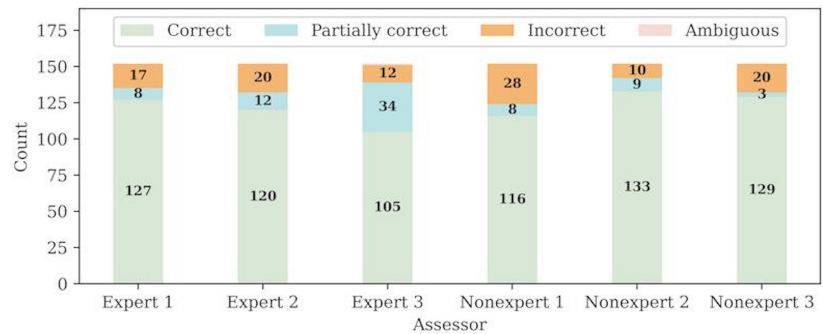

**Table 3.** Experiment 2: ChatGPT robustness and ablation study analysis. Overall response categorization between nonexpert and expert assessors. The top left section indicates the number of instances, regardless of correctness, where all nonexpert and expert assessors categorized a question identically. The bottom right section indicates the total number of questions categorized as correct by all nonexpert and expert assessors.

|  | Expert | | | |
|---|---|---|---|---|
|  | Identical | Mismatch | Correct | Other |
| Nonexpert | | | | |
| Identical | 95 | 27 | —[a] | — |
| Mismatch | 9 | 21 | — | — |
| Correct | — | — | 92 | 22 |
| Other | — | — | 4 | 34 |

[a]Not applicable.

# Discussion

## Principal Results

The hypothesis that ChatGPT is currently unsuitable for self-diagnosis is proved. From each of the 3 addressed assessments with a passing threshold of 60%, GPT-4.0 marginally only passed one. The analysis indicates that GPT-4.0 is generally unclear and incorrect when providing medical information. However, when GPT-4.0 provides correct responses, it remains robust enough to continue answering these questions accurately even when some information is missing. Table 4 summarizes the results of the evaluation procedure for GPT-4.0.

**Table 4.** Results summarization of the evaluation procedure for GPT-4.0. With a minimum threshold of 60%, the GPT-4.0 responses are deemed generally unclear and incorrect, while exhibiting robustness when providing correct answers.

|  | Definition | Success rate (%) | Pass or fail |
|---|---|---|---|
| Are the responses clear? | Agreement reached between all assessors | 34 | Fail |
| Are the responses genuinely correct? | Responses categorized as correct by all assessors | 31 | Fail |
| Are the responses robust? | Responses continue to be categorized as correct by all assessors | 61 | Pass |

## Implications

Recognizing that the GPT-4.0 responses are often ambiguous, its medical advice should be accepted with caution since these responses can vary widely in interpretation depending on the individual. Additionally, since only a small number of responses were found to be genuinely correct, LLMs still require improvement before they can reliably be used in a medical setting. Without these improvements, LLMs risk misinforming individuals.

Even though ChatGPT cannot currently be used for self-diagnosis, substantial evidence indicates that LLMs are continuously improving, suggesting their potential future use in health care systems. As EvalPrompt was also conducted on an earlier GPT-3.5 version, as detailed in Multimedia Appendix 3, the significant difference between the model performances proves that LLMs continue to improve as time progresses.

With ChatGPT being highly acclaimed for its success in passing medical examinations, researchers have proposed using ChatGPT in areas such as medical education and medical report creation [17,20]. However, ChatGPT's ability to answer examination questions does not inherently equate to genuine medical comprehension and proficiency. Instead, using ChatGPT in these medical settings can undermine health care systems since ChatGPT's overconfidence can result in misinforming individuals.

Unlike practicing clinicians, ChatGPT lacks formal testing and accreditation for its abilities. It has not undergone accredited medical education or licensing, lacks approval for clinical practice, and has not demonstrated the necessary understanding or skill set to support its claims. While clinicians face severe consequences for errors, such as medical malpractice charges or license revocation, ChatGPT lacks such liability. Thus, relying on ChatGPT before it becomes accurate, consistent, and robust poses a significant risk of misleading health care practitioners and the general public.

### Limitations

This study has signified a substantial advancement in the quality of medical advice that LLMs can provide and their potential utility in the health care industry. Although the USMLE questions provide a solid foundation for simulating the self-diagnosis process via LLMs, the prompts contain extensive detail. In other words, an ordinary individual typically would not input paragraphs of information when self-diagnosing and may lack explicit technical health care knowledge. Therefore, future research can explore using a dataset of self-diagnosis questions that contains less detailed information to accurately assess the capability of LLMs. As a result, developing larger and more realistic self-diagnosing datasets can enhance the training of LLMs, in turn improving its performance. Additionally, LLMs are inclined to excel in tasks well-represented in the training data, potentially leading to lower performance in niche problems [17]. This raises equity concerns, as questions concerning underrepresented groups may endure poor performance. Although not explicitly examined in this study, future research could generate datasets with diverse question designs across medical specializations to ensure all fields are represented in training.

### Conclusions

While LLMs make headlines for passing medical licensing examinations and are consequently being considered as candidates to train the next generation of health care professionals, it is evident that LLMs' capabilities are (understandably) modest at this time. More importantly, this misplaced trust in these systems can lead to reliance on their use for self-diagnosis by the public. In constructing EvalPrompt for assessing the capabilities of LLMs in medical contexts, the extent of misinformation inversely correlates to the LLMs' performance. LLMs that offer unclear and inaccurate responses are more likely to misinform individuals. Ultimately, this investigation presents challenges for machine learning researchers to build more transparent artificial intelligence–powered models capable of reasoning and responding responsibly, while also highlighting the need for a dataset that incorporates more realistic prompt styles.

### Acknowledgments
Sirisha Rambhatla, Ph.D., would like to acknowledge the support of the Natural Sciences and Engineering Research Council of Canada (NSERC) Discovery Grant (RGPIN-2022-03512).

### Data Availability
The step 1 United States Medical Licensing Exam dataset along with the GPT-4.0 answers analyzed during this study are available in the EvalPrompt repository [53]. The datasets for assessor categorizations analyzed during this study are not publicly available due to ChatGPT only being analyzed in aggregate, but are available from the corresponding author on reasonable request.

### Conflicts of Interest
VB is supported by an Academic Scholar Award from the University of Toronto Department of Psychiatry and has received research funding from the Canadian Institutes of Health Research, Brain & Behavior Foundation, Ontario Ministry of Health Innovation Funds, Royal College of Physicians and Surgeons of Canada, Department of National Defence (Government of Canada), New Frontiers in Research Fund, Associated Medical Services Inc Healthcare, American Foundation for Suicide Prevention, Roche Canada, Novartis, and Eisai.

### Multimedia Appendix 1
The selected ChatGPT-4.0 model along with the corresponding parameters.
[DOCX File (Microsoft Word File), 25 KB-Multimedia Appendix 1]

### Multimedia Appendix 2
Procedures and guidelines were provided to the assessors to evaluate the LLM responses. LLM: large language model.
[DOCX File (Microsoft Word File), 143 KB-Multimedia Appendix 2]

### Multimedia Appendix 3
Categorization results for ChatGPT-3.5 in comparison to ChatGPT-4.0.
[DOCX File (Microsoft Word File), 414 KB-Multimedia Appendix 3]

## Abbreviations

**A:** ambiguous
**C:** correct
**EvalPrompt:** evaluation of large language model prompts
**I:** incorrect
**LLM:** large language model
**PC:** partially correct
**USMLE:** United States Medical Licensing Exam